\documentclass[useAMS,usenatbib]{mn2e}
\usepackage{epsf,graphicx,rotating,url}

\title[Near-IR black hole mass estimates in AGN]{A near-infrared relationship for estimating black hole masses in active galactic nuclei}

\author[H. Landt et al.]{Hermine Landt$^{1,2}$\thanks{E-mail: hermine.landt@durham.ac.uk}, Martin J. Ward$^1$, Bradley M. Peterson$^{3,4}$, Misty C. Bentz$^5$, 
\newauthor
Martin Elvis$^6$, Kirk T. Korista$^7$ and Margarita Karovska$^6$ \\ 
$^1$Department of Physics, Durham University, South Road, Durham, DH1 3LE \\ 
$^2$School of Physics, University of Melbourne, Parkville, VIC 3010, Australia \\ 
$^3$Department of Astronomy, The Ohio State University, 
140 West 18th Avenue, Columbus, OH 43210, USA \\
$^4$Center for Cosmology and AstroParticle Physics, The Ohio State University, 
191 West Woodruff Avenue, Columbus, OH 43210, USA \\
$^5$ Department of Physics and Astronomy, Georgia State University, 25 Park Place Office 610, Atlanta, GA 30303, USA \\
$^6$Harvard-Smithsonian Center for Astrophysics, 60 Garden Street, 
Cambridge, MA 02138, USA \\
$^7$Department of Physics, Western Michigan University, 
1903 W. Michigan Avenue, Kalamazoo, MI 49008, USA}

\begin{document}

\def\la{\mathrel{\hbox{\rlap{\hbox{\lower4pt\hbox{$\sim$}}}\hbox{$<$}}}}
\def\ga{\mathrel{\hbox{\rlap{\hbox{\lower4pt\hbox{$\sim$}}}\hbox{$>$}}}}

\font\sevenrm=cmr7
\def\OIII{[O~{\sevenrm III}]}
\def\FeII{Fe~{\sevenrm II}}
\def\FeIIf{[Fe~{\sevenrm II}]}
\def\SIII{[S~{\sevenrm III}]}
\def\HeI{He~{\sevenrm I}}
\def\HeII{He~{\sevenrm II}}
\def\NeV{[Ne~{\sevenrm V}]}
\def\OIV{[O~{\sevenrm IV}]}

\def\cloudy{{\sevenrm CLOUDY}}

\date{Accepted ~~. Received ~~; in original form ~~}

\pagerange{\pageref{firstpage}--\pageref{lastpage}} \pubyear{2013}

\maketitle

\label{firstpage}

\begin{abstract}

Black hole masses for samples of active galactic nuclei (AGN) are
currently estimated from single-epoch optical spectra using scaling
relations anchored in reverberation mapping results. In particular,
the two quantities needed for calculating black hole masses, namely,
the velocity and the radial distance of the orbiting gas are derived
from the widths of the Balmer hydrogen broad emission lines and the
optical continuum luminosity, respectively. We have recently presented
a near-infrared (near-IR) relationship for estimating AGN black hole
masses based on the widths of the Paschen hydrogen broad emission
lines and the total 1~$\mu$m continuum luminosity. The near-IR offers
several advantages over the optical: it suffers less from dust
extinction, the AGN continuum is observed only weakly contaminated by
the host galaxy and the strongest Paschen broad emission lines
Pa$\alpha$ and Pa$\beta$ are unblended. Here we improve the
calibration of the near-IR black hole mass relationship by increasing
the sample from 14 to 23 reverberation-mapped AGN using additional
spectroscopy obtained with the Gemini Near-Infrared Spectrograph
(GNIRS). The additional sample improves the number statistics in
particular at the high luminosity end.

\end{abstract}


\section{Introduction}

The discovery of tight correlations between a galaxy's central black
hole mass and the luminosity, velocity dispersion and mass of its
stellar bulge \citep{Mag98, Geb00, Ferr00, Marc03, Haer04} is expected
to strongly constrain how galaxies form and grow over cosmic
time. Therefore, much effort in particular goes into measuring the
rate of black hole growth \citep[e.g.,][]{Yu02, Heck04,
  Kelly09}. Since this requires both large samples of galaxies with
easily obtainable black hole mass estimates and sources that probe the
highest redshifts, such studies rely heavily on active galactic nuclei
(AGN).

In general, the masses of black holes at the centres of active
galaxies are measured based on the gravitational force they exert on
other massive objects such as stars or gas clouds. But, whereas
measurements of the stellar dispersion require spectroscopy at very
high angular resolution in order to resolve the black hole's sphere of
influence, the velocity dispersion of the broad-emission line gas
present at the centre of AGN can be readily measured from a single
long-slit optical, ultraviolet (UV) or near-infrared (near-IR)
spectrum. Then, assuming that the dynamics of the broad-emission line
gas is dominated by gravitational forces, one can use the virial
theorem to calculate black hole masses:

\begin{equation}
\label{virial}
M_{\rm BH} = f \frac{R \Delta V^2}{G},
\end{equation}

\noindent
where $R$ is the radial distance of the broad emission line gas from
the black hole, $\Delta V$ is the velocity dispersion of the gas, $G$
is the gravitational constant and $f$ is a scaling factor that depends
on the (unknown) dynamics and geometry of the broad line region (BLR).

The BLR radius can be directly measured through reverberation mapping,
a technique which determines the light-travel time-delayed lag with
which the flux of the BLR responds to changes in the {\it ionising}
continuum flux \citep[see reviews by][]{Pet93, Netz97,
  Pet06}. However, since reverberation mapping campaigns are observing
time intensive, at present the BLR radii of fewer than 50 AGN have
been measured \citep{Kaspi00, Pet04, Kaspi07, Bentz08, BentzLick,
  Barth11}. Therefore, the so-called radius-luminosity ($R$--$L$)
relationship is used to estimate the BLR sizes for large samples of
AGN. As \citet{Pet93}, \citet{Wan99}, \citet{Kaspi00} and
\citet{Bentz09} have shown, the BLR lags obtained from reverberation
mapping campaigns correlate with the optical luminosity (of the
ionising component) largely as expected from simple photoionisation
arguments. The continuum luminosity $L$ is usually measured in the
optical (at rest-frame 5100~\AA) or in the UV.

Given the importance of the $R$--$L$ relationship for AGN black hole
mass determinations, alternatives to the optical and UV continuum
luminosity are being explored, e.g., the X-ray luminosity and broad
H$\beta$ line luminosity \citep{Kaspi05, Greene10}, the \OIII~$\lambda
5007$~\AA~emission-line luminosity \citep{Greene10}, the Pa$\alpha$
and Pa$\beta$ emission-line luminosities \citep{Kim10}, and the
mid-infrared \NeV~$\lambda14.32~\mu$m and \OIV~$\lambda 25.89~\mu$m
emission-line luminosities \citep{Das08, Greene10}. In \citet{L11b} we
proposed the 1~$\mu$m continuum luminosity as an efficient alternative
that connects directly to the spectrum of the ionising source. Then,
in \citet{L11a} we introduced the relationship between the black hole
mass and the near-IR virial product, i.e., the product between the
total 1~$\mu$m continuum luminosity and the width of the Paschen
hydrogen broad emission lines.

The near-IR potentially affords some advantages over the optical and
UV. First, the observed optical continuum can be severely contaminated
by host galaxy starlight if a large slit is used (as is often the case
in reverberation studies), especially in low-redshift sources that
have a weak AGN or sources with a luminous stellar bulge, and so does
not give directly the ionising flux. High-resolution, deep optical
images are then required to estimate the host galaxy starlight
enclosed in the aperture in order to correct the optical spectra
\citep{Bentz06a, Bentz09}. Secondly, the optical hydrogen
broad-emission lines, in particular H$\beta$, and most of the UV broad
emission lines are strongly blended with other species, making it
difficult to reliably measure the line width. Thirdly, all optical and
UV measurements may suffer from dust extinction.

Near-IR spectroscopy is now available at excellent observing sites
that regularly achieve subarcsecond seeing. Additionally, the seeing
is improved at longer wavelengths (e.g., by a factor of $\sim 1.3$ at
1~$\mu$m relative to 5100~\AA). This means that the host galaxy flux
contribution in the near-IR can be reduced to a minimum. Note,
however, that for the same (small) aperture the optical spectrum would
be less dominated by host galaxy light than the near-IR one since the
AGN to host galaxy flux ratio is intrinsically higher at optical
wavelengths. Then, as we have shown in \citet{L08a}, the strongest
Paschen hydrogen broad emission lines Pa$\alpha$ and Pa$\beta$ are
observed to be free from strong blends offering the opportunity to
reliably measure their widths. Furthermore, since attenuation by dust
is much reduced in the near-IR compared to the optical and UV, our
near-IR relationship should be particularly useful as an alternative
for dust-obscured AGN.

Here we present an improved calibration of our near-IR relationship
for estimating black hole masses in AGN. The paper is structured as
follows. In Section 2, we introduce the enlarged sample that includes
additional sources, which were observed as described in Section 3. In
Section 4, we present an improved calibration of the near-IR $R$--$L$
relationship, and in Section 5, we derive the revised relationship
between the black hole mass and the near-IR virial product. A few
sources in our sample are found to have near-IR luminosities dominated
by host galaxy starlight and we discuss them in the context of the
near-IR relationships in Section 6. In Section 7, we summarize our
results and present our conclusions. Throughout this paper we have
assumed cosmological parameters $H_0 = 70$ km s$^{-1}$ Mpc$^{-1}$,
$\Omega_{\rm M}=0.3$, and $\Omega_{\Lambda}=0.7$.

\section{The sample}

\begin{table*}
\begin{turn}{90}
\begin{minipage}{230mm}
\vspace*{-0.5cm}
\caption{\label{sample} 
General properties of the additional sample}
\begin{tabular}{lccccccccccccr}
\hline
Object name & R.A. (J2000.0) & Decl. (J2000.0) & z & J & $R_{\rm H\beta}$ & Ref. & $M_{\rm BH}$ & Ref. & log $\nu L_{\rm 1\mu m}$ & line & type & FWHM & $\sigma_{\rm line}$ \\
&&&& (mag) & (lt-days) && ($10^6$ $M_{\odot}$) && (erg s$^{-1}$) &&& (km s$^{-1}$) & (km s$^{-1}$) \\
(1) & (2) & (3) & (4) & (5) & (6) & (7) & (8) & (9) & (10) & (11) & (12) & (13) & (14) \\
\hline
PG 0026$+$129 & 00 29 13.60 & $+$13 16 03.0 & 0.145$^\star$&13.58 & $111_{-28}^{+24}$ & B09 & 393$\pm$96  & P04 & 44.72 & Pa$\alpha$ & e    & 1793 & 1798 \\
PG 0052$+$251 & 00 54 52.10 & $+$25 25 38.0 & 0.155 & 13.89 & $90_{-24}^{+25}$  & B09 & 369$\pm$76          & P04 & 44.46 & Pa$\alpha$ & i    & 4133 & 1750 \\ 
3C 120        & 04 33 11.10 & $+$05 21 15.6 & 0.033 & 12.67 & $27.6_{-0.9}^{+1.1}$ & G12 & 66$\pm$7         & G12 & 43.45 & Pa$\beta$  & i    & 2733 & 1815 \\
PG 0804$+$761 & 08 10 58.60 & $+$76 02 42.5 & 0.100 & 12.98 & 147$\pm$19        & B09 & 693$\pm$83          & P04 & 44.58 & Pa$\alpha$ & lack & 2304 & 1193 \\
NGC 3516      & 11 06 47.49 & $+$72 34 06.9 & 0.009 & 11.13 & $12_{-2}^{+1}$    & D10 & $32_{-4}^{+3}$      & D10 & 43.04 & Pa$\beta$ & lack & 4469 & 1527 \\
NGC 4051      & 12 03 09.61 & $+$44 31 52.8 & 0.002 & 11.65 & 1.9$\pm$0.5       & D10 & $1.7_{-0.5}^{+0.6}$ & D10 & 41.51 & Pa$\beta$ & e & 1681 & 762 \\
PG 1211$+$143 & 12 14 17.70 & $+$14 03 12.6 & 0.081 & 13.15 & $94_{-42}^{+26}$  & B09 & 146$\pm$44          & P04 & 44.33 & Pa$\alpha$ & e    & 1601 &  955 \\
PG 1307$+$085 & 13 09 47.00 & $+$08 19 48.2 & 0.155 & 14.16 & $106_{-47}^{+36}$ & B09 & 440$\pm$123         & P04 & 44.61 & Pa$\alpha$ & i    & 2982 & 1631 \\
Mrk 279       & 13 53 03.45 & $+$69 18 29.6 & 0.030 & 12.19 & 17$\pm$4          & B09 & 35$\pm$9            & P04 & 43.43 & Pa$\beta$  & i    & 3568 & 1580 \\
3C 390.3      & 18 42 09.00 & $+$79 46 17.1 & 0.056 & 13.60 & $24_{-7}^{+6}$    & B09 & 287$\pm$64          & P04 & 44.03 & Pa$\alpha$ & i    & 7460 & 2612 \\
Mrk 1513      & 21 32 27.81 & $+$10 08 19.5 & 0.063 & 12.84 & $13.3_{-0.5}^{+0.6}$ & G12 & 41$\pm$9         & G12 & 44.08 & Pa$\alpha$ & e    & 1905 & 1265 \\
\hline
\end{tabular}

\parbox[]{21.5cm}{The columns are: (1) object name; (2) and (3)
  position, and (4) redshift from the NASA/IPAC Extragalactic Database
  (NED); (5) $J$ band magnitude from the Two Micron All-Sky Survey
  \citep[2MASS;][]{Cutri03}; (6) radius of the H$\beta$ broad-emission
  line region (in light-days); (7) reference for the H$\beta$ radius,
  where B09: \citet{Bentz09}, D10: \citet{Denney10}, G12:
  \citet{Grier12}; (8) black hole mass (in solar masses) calculated
  from the line dispersion assuming a scaling factor of $f=5.5$; (9)
  reference for the black hole mass, where D10: \citet{Denney10}, G12:
  \citet{Grier12}, P04: \citet{Pet04}; (10) integrated total 1-$\mu$m
  continuum luminosity; (11) Paschen emission line with the highest
  signal-to-noise ratio; (12) type of transition between the broad and
  narrow emission line component, where i: inflected (i.e., transition
  point is obsvious), e: estimated (i.e., transition point is
  estimated) and lack: absent narrow emission line component; (13)
  full width at half maximum (FWHM) and (14) dispersion of the broad
  emission line component (both uncorrected for instrumental
  resolution). Errors represent $1\sigma$ uncertainties.

\smallskip

\parbox[]{21.5cm}{$^\star$ We measure a redshift of $z=0.145$ from
  narrow emission lines, whereas a value of $z=0.142$ is listed in
  NED.}

}

\end{minipage}
\end{turn}
\end{table*}

\begin{figure}
\centerline{
\includegraphics[scale=0.4]{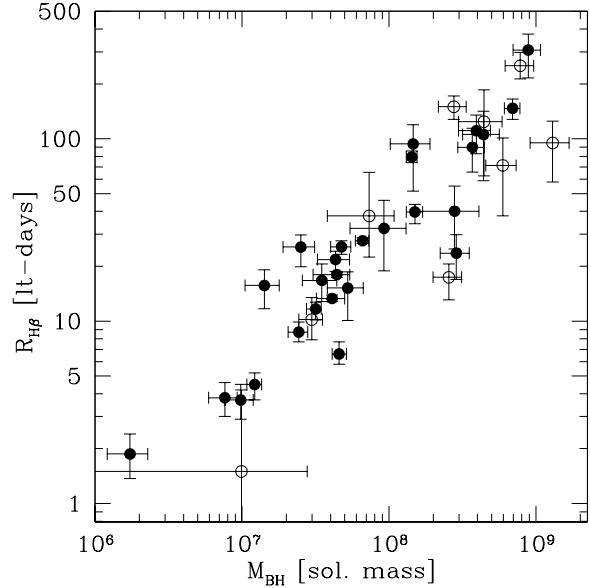}
}
\caption{\label{totsample} The radius of the H$\beta$ broad-emission
  line region (in light-days) versus the black hole mass (in solar
  masses) for the reverberation-mapped AGN sample with (filled
  circles) and without near-IR spectroscopy (open circles).}
\end{figure}

\begin{figure*}
\centerline{
\includegraphics[bb=18 335 591 718,clip=true,scale=0.9]{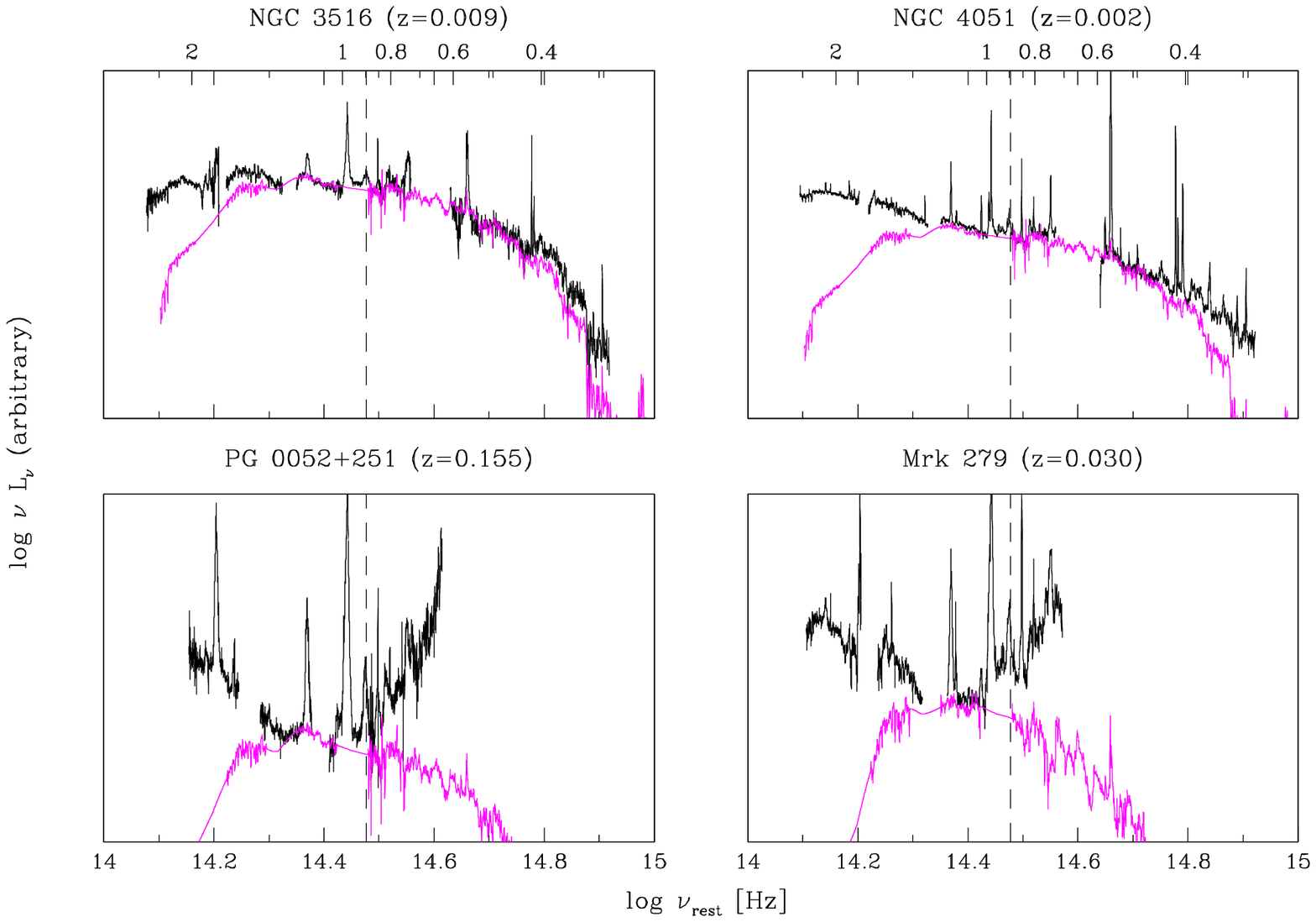}
}
\caption{\label{SEDgal} Rest-frame spectral energy distributions
  (SEDs) for the two sources (NGC~3516 and NGC~4051) that are strongly
  affected by host galaxy light (top panels) and for the two sources
  (PG~0052$+$251 and Mrk~279) that have an SED dominated by AGN
  emission although estimates based an {\sl HST} image decomposition
  give relatively high host galaxy flux contributions (bottom
  panels). The near-IR spectra are normalized at 1 $\mu$m (dashed
  line) and non-simultaneous optical spectra from \citet{Ken92} and
  \citet{Mous06} are also shown for NGC~3516 and NGC~4051. The host
  galaxy templates from \citet{Man01} are overplotted
  (magenta). Wavelength units in $\mu$m are labeled on the top axis.}
\end{figure*}

\citet{Pet04} reanalysed the available reverberation mapping data of a
sample of 37 AGN and presented improved black hole mass determinations
for 35 objects. Later, \citet{Bentz06b}, \citet{Denney06},
\citet{Denney10} and \citet{Grier12} refined the black hole mass
estimates for some of these sources and added a new determination for
the source Mrk~290.

Our original sample observed with the SpeX spectrograph \citep{Ray03}
on the NASA Infrared Telescope Facility (IRTF), a 3 m telescope on
Mauna Kea, Hawai'i, included 16/36 reverberation-mapped AGN from the
Peterson et al. sample \citep{L08a}. For observations with the Gemini
Near-Infrared Spectrograph (GNIRS) at the Gemini North observatory, an
8.1~m telescope on Mauna Kea, Hawai'i, we have selected all the
remaining northern ($\delta > -37^\circ$) reverberation-mapped AGN. We
were granted observations of one source (3C~120) in the Science
Verification (SV) phase of GNIRS and of eight sources in semester
2011B. Additionally, two reverberation-mapped AGN (NGC~4051 and
Mrk~279) were observed with the IRTF by \citet{Rif06} and we made use
of their publicly available spectra. Therefore, our enlarged sample
contains 27/36 reverberation-mapped AGN from the Peterson et
al. sample. Table \ref{sample} lists the general properties of the
additional sample.

Fig. \ref{totsample} shows the radius of the H$\beta$ broad-emission
line region ($R_{\rm H\beta}$) versus the black hole mass for the
reverberation-mapped AGN with (filled circles) and without near-IR
spectroscopy (open circles). The remaining sample of nine sources
would increase the number statistics in particular at the high end of
values. Therefore, obtaining near-IR spectroscopy of them could lead
to a significant improvement of the calibration of our near-IR
relationship for estimating black hole masses.

Four sources in our enlarged sample, namely, Mrk~590, NGC~3227,
NGC~3516 and NGC~4051, were found to be in a very low luminosity state
with a continuum spectral energy distribution (SED) strongly dominated
by host galaxy emission (see Fig. \ref{SEDgal}, top panels, and
\citet{L11a}, their Fig. 6). We have excluded these sources from the
calibration of the near-IR relationships in Sections \ref{radiuslum}
and \ref{nearIRvirial}, but we discuss them in the general context in
Section \ref{lowlum}, since an upper limit to the integrated 1~$\mu$m
continuum luminosity of the AGN can be obtained from the total
observed value.

\section{The observations}

\begin{table*}
\caption{\label{obslog} 
Gemini Journal of observations}
\begin{tabular}{llccrrrlrc}
\hline
Object name & observation & exposure & airmass & \multicolumn{3}{c}{continuum S/N} 
& \multicolumn{3}{c}{telluric standard star} \\
& date & (sec) && J & H & K & name & distance & airmass \\
&&&&&&&& ($^{\circ}$) & \\
(1) & (2) & (3) & (4) & (5) & (6) & (7) & (8) & (9) & (10) \\
\hline
PG 0026$+$129 & 2011 Aug 3  & 8$\times$120 & 1.144 &  62 &  95 &  69 & HD222558 & 12.1 & 1.113 \\
PG 0052$+$251 & 2011 Aug 3  & 8$\times$120 & 1.012 &  69 &  89 & 119 & HD222558 & 23.1 & 1.149 \\ 
3C 120        & 2010 Dec 15 & 2$\times$120 & 1.046 &  49 &  55 & 164 & HD34035  & 10.7 & 1.053 \\
PG 0804$+$761 & 2011 Nov 30 & 8$\times$120 & 1.828 &  84 & 106 & 160 & HD48049  &  8.8 & 1.786 \\
NGC 3516      & 2011 Dec 8  & 4$\times$120 & 1.791 & 130 & 168 & 149 & HD48049  & 20.2 & 1.606 \\
PG 1211$+$143 & 2011 Dec 18 & 8$\times$120 & 1.473 & 147 &  70 & 119 & HD101060 &  8.9 & 1.402 \\
PG 1307$+$085 & 2011 Aug 11 & 8$\times$120 & 1.851 &  50 &  48 &  92 & HD116960 &  5.5 & 1.789 \\
3C 390.3      & 2011 Aug 4  & 8$\times$120 & 2.034 & 132 & 156 & 187 & HIP942   & 27.2 & 1.901 \\
Mrk 1513      & 2011 Aug 3  & 8$\times$120 & 1.034 & 132 &  92 &  85 & HD210501 & 11.9 & 1.116 \\
\hline
\end{tabular}

\parbox[]{14.3cm}{The columns are: (1) object name; (2) observation date; (3)
  exposure time; (4) average airmass; S/N in the continuum over $\sim
  100$~\AA~measured at the central wavelength of the (5) J, (6) H, and
  (7) K band; for the star used to correct for telluric absorption (8) name, 
  (9) distance from the source, and (10) average airmass.}

\end{table*}

We observed in queue mode with the Gemini Near-Infrared Spectrograph
\citep[GNIRS;][]{gnirs}, which has recently been recommissioned at
the~Gemini North observatory. One source (3C~120) was observed
during the Science Verification (SV) phase of GNIRS in semester 2010B
(Program ID: GN-2010B-SV-165) and eight sources in semester 2011B
(Program ID: GN-2011B-Q-97). We used the cross-dispersed mode with the
short camera at the lowest resolution (31.7~l~mm$^{-1}$ grating), thus
covering the entire wavelength range of $0.9-2.5$~$\mu$m without
inter-order contamination. We chose a slit of $0.3''\times7''$ for the
source 3C~120 and, due to a poor weather allocation, a larger slit of
$0.675''\times7''$ for the sources observed in semester 2011B. This
set-up gives an average spectral resolution of full with at
half-maximum (FWHM) $\sim 180$~km~s$^{-1}$ for 3C~120 and of FWHM$\sim
400$~km~s$^{-1}$ for the remainder of the sources. The chosen exposure
times ensured that we obtained spectra with a high signal-to-noise
ratio (S/N $\sim 100$) in order to reliably measure the broad emission
line profiles.

Either before or after each source, we observed a nearby (in position
and air mass) A0~V star with accurate near-IR magnitudes. These stars
were used to correct the source spectra for telluric absorption and
for flux calibration. Flats and arcs were taken with each
source/telluric standard star pair. In Table \ref{obslog} we list the
journal of observations. The data were reduced using the Gemini/IRAF
package (version 1.11) with GNIRS specific tools
\citep{gnirssoft}. The data reduction steps included preparation of
calibration and science frames, processing and extraction of spectra
from science frames, wavelength calibration of spectra, telluric
correction and flux-calibration of spectra, and merging of the
different orders into a single, continuous spectrum. The spectral
extraction width was adjusted interactively for each telluric standard
star and source to include all the flux in the spectral trace. The
final spectra were corrected for Galactic extinction using the IRAF
task \mbox{\sl onedspec.deredden} with input $A_{\rm V}$ values
derived from Galactic hydrogen column densities published by
\citet{DL90}. The results are shown in Fig. \ref{gnirsspec}.

\section{The near-IR $R$--$L$ relationship} \label{radiuslum}

\begin{table*}
\caption{\label{hostir} 
Estimates of the near-IR host galaxy contribution}
\begin{tabular}{lcllccccc}
\hline
Object Name & A$_{(1+z)5100}$ & host & ref. & \multicolumn{2}{c}{near-IR spectrum} & {\sl HST} flux & 
log $\nu L_{\rm 1\mu m}$ & AGN/ \\
& (mag) & type && aperture & PA & ($(1+z)5100$~\AA) & (erg s$^{-1}$) & host \\
&&&& (arcsec$^2$) & ($^\circ$) & (erg s$^{-1}$ cm$^{-2}$ \AA$^{-1}$) \\
(1) & (2) & (3) & (4) & (5) & (6) & (7) & (8) & (9) \\
\hline
PG 0026$+$129 & 0.105 & E    & B09 & 0.7 $\times$ 5.8 & 0   & 7.65E$-$17 & 43.70 &  9 \\
PG 0052$+$251 & 0.115 & Sb   & H02 & 0.7 $\times$ 4.3 & 0   & 2.67E$-$16 & 44.22 &  0.7 \\ 
3C 120        & 0.509 & S0   & P07 & 0.3 $\times$ 1.9 & 0   & 7.30E$-$17 & 42.26 & 14 \\
PG 0804$+$761 & 0.014 & E    & B09 & 0.7 $\times$ 5.5 & 0   & 2.42E$-$17 & 42.83 & 55 \\
NGC 3516      & 0.030 & Sb   & RC3 & 0.7 $\times$ 4.5 & 0   & 3.06E$-$15 & 42.66 &  1 \\
NGC 4051      & 0.000 & Sb   & RC3 & 0.8 $\times$ 1.6 & 132 & 2.04E$-$15 & 41.17 &  1 \\
PG 1211$+$143 & 0.000 & E    & B09 & 0.7 $\times$ 5.5 & 0   & 1.63E$-$16 & 43.46 &  6 \\
PG 1307$+$085 & 0.000 & E    & B09 & 0.7 $\times$ 3.7 & 0   & 2.94E$-$17 & 43.35 & 17 \\
Mrk 279       & 0.000 & Sa   & P07 & 0.8 $\times$ 1.7 & 0   & 1.12E$-$15 & 43.33 &  0.3 \\
3C 390.3      & 0.085 & Sa   & B09 & 0.7 $\times$ 5.6 & 0   & 3.58E$-$16 & 43.40 &  3 \\
Mrk 1513      & 0.114 & S0/a & P07 & 0.7 $\times$ 5.7 & 0   & 1.68E$-$16 & 43.22 &  6 \\
\hline
\end{tabular}

\parbox[]{14.5cm}{The columns are: (1) object name; (2) Galactic
  extinction at rest-frame 5100~\AA; (3) Hubble type of the host
  galaxy; (4) reference for the host type, where H02: \citet{Hamil02},
  P07: \citet{Petr07}, B09: \citet{Bentz09}, RC3: \citet{RC3}; (5)
  near-IR extraction aperture; (6) near-IR slit position angle, where
  PA=0$^{\circ}$ corresponds to E-W orientation and is defined E
  through N; (7) host galaxy flux at rest-frame 5100~\AA, corrected
  for Galactic extinction using the values in column (3); (8)
  integrated host galaxy 1-$\mu$m continuum luminosity; and (9)
  luminosity ratio between AGN and host galaxy at rest-frame
  1-$\mu$m.}

\end{table*}

\begin{figure}
\centerline{
\includegraphics[scale=0.4]{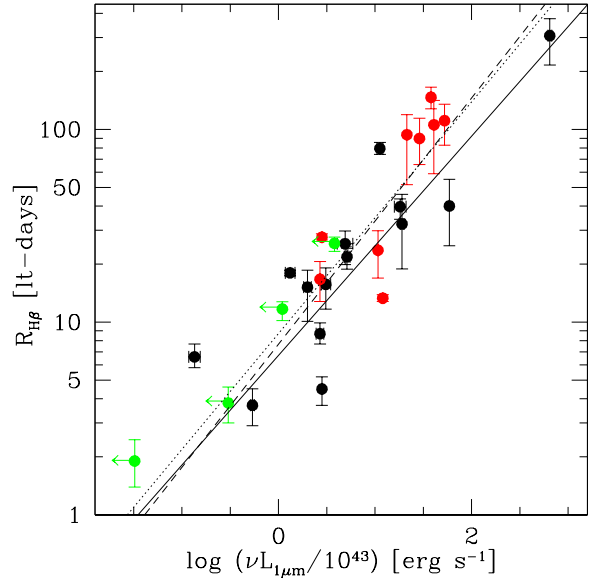}
}
\caption{\label{lag} The radius of the H$\beta$ broad-emission line
  region (in light-days) versus the integrated total 1~$\mu$m
  continuum luminosity. Black and red filled circles indicate the
  original sample of \citet{L11b} and our additional sample,
  respectively. The sources with a continuum SED dominated by host
  galaxy emission are shown as green filled circles. The dashed, solid
  and dotted lines indicate the correlations obtained with the BCES,
  FITEXY and GaussFit routines, respectively.}
\end{figure}

\begin{table*}
\begin{center}
\caption{\label{fits} 
Best-fits for the relation $\log R_{\rm H\beta} = K + \alpha\log(\nu L_{\nu}/10^{43})$}
\begin{tabular}{lcccccc}
\hline
& \multicolumn{3}{c}{original sample (14 obj.)} & \multicolumn{3}{c}{enlarged sample (23 obj.)} \\
\hline
Type & K & $\alpha$ & S$^{\star}$ & K & $\alpha$ & S$^{\star}$ \\
\hline
\multicolumn{7}{c}{BCES} \\
\hline
Avg & 0.91$\pm$0.11 & 0.56$\pm$0.10          & & 0.89$\pm$0.11 & 0.64$\pm$0.09 & \\
MC  & 0.90$\pm$0.13 & $0.56^{+0.12}_{-0.14}$ & & 0.88$\pm$0.12 & 0.64$\pm$0.10 & \\
\hline
\multicolumn{7}{c}{FITEXY} \\
\hline
Avg & 0.82$\pm$0.10 & 0.52$\pm$0.10 & 53       & 0.83$\pm$0.09 & 0.57$\pm$0.08 & 53 \\
MC  & 0.81$\pm$0.11 & 0.53$\pm$0.11 & 52$\pm$2 & 0.81$\pm$0.10 & 0.58$\pm$0.09 & 52$\pm$1 \\
\hline
\multicolumn{7}{c}{GaussFit} \\
\hline
Avg & 1.00$\pm$0.10 & 0.54$\pm$0.10 &          & 0.94$\pm$0.09 & 0.60$\pm$0.09 & \\
MC  & 0.95$\pm$0.14 & 0.56$\pm$0.12 &          & $0.91^{+0.12}_{-0.13}$ & 0.62$\pm$0.11 & \\
\hline
\end{tabular}
\end{center}

\parbox[]{10cm}{$^{\star}$ Scatter calculated as the percentage of
  the $\log R_{H\beta}$ value that, when added in quadrature to the
  error value, gives $\chi^2_{\nu}=1$.}

\end{table*}

In \citet{L11a} we have shown that the accretion disc spectrum of AGN,
which is believed to be the main source of ionising radiation, extends
well into the near-IR and still dominates the continuum at $\sim
1$~$\mu$m. Therefore, a single-epoch near-IR spectrum of a
broad-emission line AGN can in principle be used to estimate the BLR
radius. We verified this conjecture in \citet{L11b}, where we
introduced the near-IR radius -- luminosity ($R$--$L$)
relationship. Here we present an improved calibration of it.

In Fig. \ref{lag} we plot the radius of the H$\beta$ broad-emission
line region ($R_{\rm H\beta}$) versus the integrated total 1~$\mu$m
continuum luminosity for our enlarged sample of reverberation-mapped
AGN (27 objects). Values for the radius of the H$\beta$ broad-line
region are from \citet{Bentz09}, \citet{Denney10} and \citet{Grier12}
(see also Table \ref{sample}). The near-IR measures for the original
sample (black filled circles) are based on our IRTF data, whereas we
have measured the integrated total 1~$\mu$m continuum luminosity for
the additional sample (red filled circles) in the GNIRS spectra
obtained at the Gemini North observatory and for Mrk~279 in the IRTF
data of \citet{Rif06}. We observed the original IRTF sample on average
twice within a period of $\sim 3$ years \citep{L08a, L11a}. For these
sources we use the mean integrated total 1~$\mu$m continuum luminosity
and the error on the mean. For the Gemini sample, which has only one
observation epoch available, we have adopted the average error of the
IRTF sample of $0.04$ dex.

The excellent atmospheric seeing on Mauna Kea, Hawai'i, where both the
IRTF and the Gemini North observatories are located, allowed us to use
a relatively narrow slit ($<1''$), which excluded most of the host
galaxy starlight. We have verified that our spectra are indeed
dominated by the AGN in two separate ways. First, we have estimated
the host galaxy contribution in the near-IR aperture using the {\sl
  Hubble Space Telescope (HST)} images of \citet{Bentz06a} and
\citet{Bentz09} and following their approach. The observed {\sl HST}
fluxes were transformed to a rest-frame wavelength of 5100~\AA~by
applying a colour correction factor based on the model bulge galaxy
template of \citet{Kin96} and were corrected for Galactic extinction
using $A_{\lambda}$ values derived from the hydrogen column densities
of \citet{DL90}. The unabsorbed rest-frame 5100~\AA~fluxes were then
used to derive rest-frame 1~$\mu$m fluxes by scaling the galaxy
template from \citet{Man01} of the appropriate Hubble type. Details of
our host galaxy flux estimates for the original sample are listed in
\citet{L11a} and for the additional sample are given in Table
\ref{hostir}. Secondly, since {\sl HST} image decomposition of AGN can
be problematic at the centre of the galaxy where the bright AGN is
degenerate with the concentration of the compact bulge, we have
checked that the shape of the observed continuum SED is dominated by
the AGN, i.e. that we observe the accretion disc and the hot dust
emission rising and falling towards longer frequencies (in a
logarithmic $\nu L_\nu$ versus $\nu$ plot), respectively, resulting in
a characteristic inflection point at the location where they meet (at
$\sim 1$ $\mu$m).

As expected, the relatively small aperture used in the near-IR leads
to relatively high AGN to host galaxy flux ratios and so we observe a
continuum SED dominated by the AGN. However, four sources (NGC~3516
and NGC~4051, see Fig. \ref{SEDgal}, top panels, and Mrk~590 and
NGC~3227, see Fig. 6 in \citet{L11a}) were found to have a continuum
SED strongly dominated by host galaxy emission. Since in these cases
the observed 1~$\mu$m continuum luminosity represents only an upper
limit for the AGN contribution, we have excluded these from the
following analysis. For two sources (PG~0052$+$251 and Mrk~279) the
decomposition of {\sl HST} images gives a relatively high host galaxy
flux contribution, however, their continuum SED is clearly dominated
by AGN emission (see Fig. \ref{SEDgal}, bottom panels).

With the additional sample (red filled circles) we have improved the
number statistics in particular at the high end of the $R$--$L$
relationship, where we previously had only two sources. More than half
of the additional AGN have radii of the H$\beta$ broad-emission line
region of $R_{\rm H\beta} \ga 50$~light-days and total 1~$\mu$m
continuum luminosities of $\nu L_{\rm 1\mu m} \ga 2 \cdot
10^{44}$~erg~s$^{-1}$.  At the low-luminosity end we find mainly
sources with a continuum dominated by host galaxy light (green filled
circles). Therefore, improving the number statistics of the $R$--$L$
relationship in this regime and extending its dynamic range to even
lower values is likely to be problematic.

We have performed linear fits of the form \mbox{$\log R_{\rm H\beta} =
  K + \alpha\log(\nu L_{\nu}/10^{43})$} following the approach of
\citet{Bentz06a} and \citet{Bentz09}. In particular, we have used the
three fitting routines BCES \citep{Akr96b}, FITEXY \citep{recipes} and
GaussFit \citep{McA94} that can incorporate errors on both variables
and, except for GaussFit, allow us to account for intrinsic
scatter. Note that accounting for intrinsic scatter has the effect of
increasing the weight given to data points with the largest errors,
which is preferred if the intrinsic dispersion is larger than the
measurement errors \citep{Tre02}. Roughly a quarter of our sample
(6/23 sources) has multiple measurements of $R_{\rm H\beta}$ and,
therefore, we have considered the following two cases: (i) using the
average derived from all measurements for a particular source and
weighted by the mean of the positive and negative errors (i.e.,
weighted averages) and (ii) using Monte-Carlo (MC) techniques to
randomly sample $R_{\rm H\beta}$ from the individual values for each
object. In Table \ref{fits} we compare our previous results from
\citet{L11b} with the results for the enlarged sample. The latter are
shown for the case of the weighted averages in Fig. \ref{lag}. Note
that the near-IR continuum luminosities and the optical broad-emission
line radii are not measured simultaneously, which may increase the
scatter in their relationship.

Simple photoionisation arguments suggest that a given emission line
will be produced at the same ionising {\it flux} in all AGN and
therefore $R \propto L^{1/2}$. We obtain with the enlarged sample in
all six cases a best-fitting slope of $\sim 0.6\pm0.1$, which is
slightly steeper than our previous results but consistent with a value
of 0.5. The errors on the slope and the intercept have not been
significantly reduced with the enlarged sample. Similarly, the scatter
stays unchanged at $\sim 50\%$, indicating that it is intrinsic rather
than observational. The scatter in the near-IR $R$--$L$ relationship
is slightly larger than that of the current optical $R$--$L$
relationship. For the entire optical sample of \citet{Bentz09} of
reverberation-mapped AGN with host galaxy-subtracted fluxes (34
sources) we found a scatter of $\sim 40\%$ \citep[][see their Table
  2]{L11b}.

\section{The near-IR virial product} \label{nearIRvirial}

\begin{figure}
\centerline{
\includegraphics[clip=true,bb=15 330 575 700,scale=0.45]{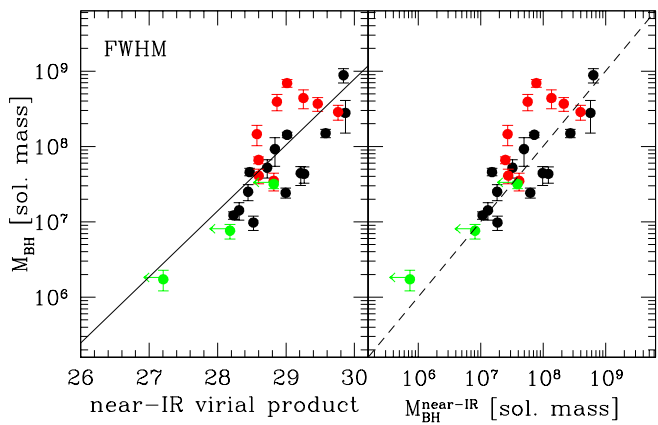}
}
\caption{\label{mbhfwhm} Black hole mass determined from optical
  reverberation campaigns versus the near-IR virial product between
  the total continuum luminosity at 1 $\mu$m and the line width of the
  Pa$\alpha$ or Pa$\beta$ broad component, whichever line had the
  higher signal-to-noise ratio (left panel) and versus the black hole
  mass calculated using the near-IR $R$--$L$ relationship and the
  virial theorem (right panel). We have used for the line width the
  full width at half maximum (FWHM). Symbols are as in
  Fig. \ref{lag}. The black solid line in the left panel indicates the
  observed correlation, whereas the black dashed line in the right
  panel shows the locus of equality.}
\end{figure}

\begin{figure}
\centerline{
\includegraphics[clip=true,bb=15 330 575 700,scale=0.45]{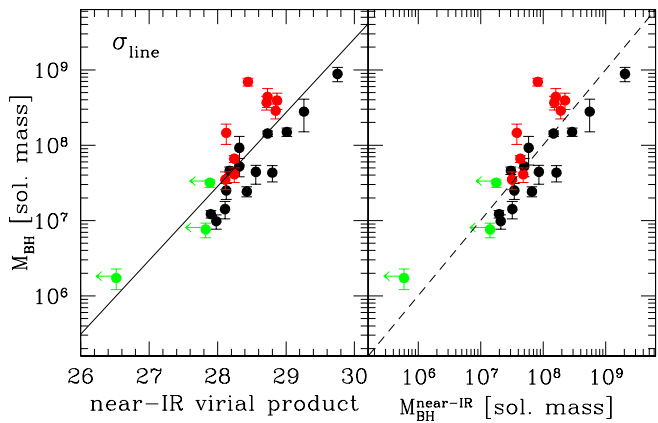}
}
\caption{\label{mbhdisp} As in Fig. \ref{mbhfwhm} using for the line
  width the line dispersion.}
\end{figure}

As we have shown in \citet{L11a}, AGN black holes masses can be
estimated from the near-IR virial product based on the 1~$\mu$m
continuum luminosity and the width of the Pa$\beta$ (or Pa$\alpha$)
broad emission line because the widths of the broad Paschen lines are
well-correlated with those of the broad Balmer lines \citep{L08a}.
The advantages of using the near-IR instead of the optical virial
product are threefold. First, since Pa$\alpha$ and Pa$\beta$ are
observed to be unblended \citep{L08a}, their width can be reliably
measured. By contrast, the H$\beta$ broad emission line is often
observed to be strongly blended with both \FeII~and
\HeII~$\lambda4686$ and often shows a ``red shelf'' most likely formed
by weak \FeII~multiplets and \HeI. Secondly, the AGN continuum around
1~$\mu$m is free from major contaminating components and can be easily
determined, unlike the optical, which can suffer from a
pseudo-continuum caused by blended and broadened
\FeII~emission. Finally, since the near-IR is much less affected by
dust extinction, it can potentially be applied to dust-obscured AGN.

\begin{figure*}
\centerline{
\includegraphics[clip=true,bb=33 172 570 705,scale=0.9]{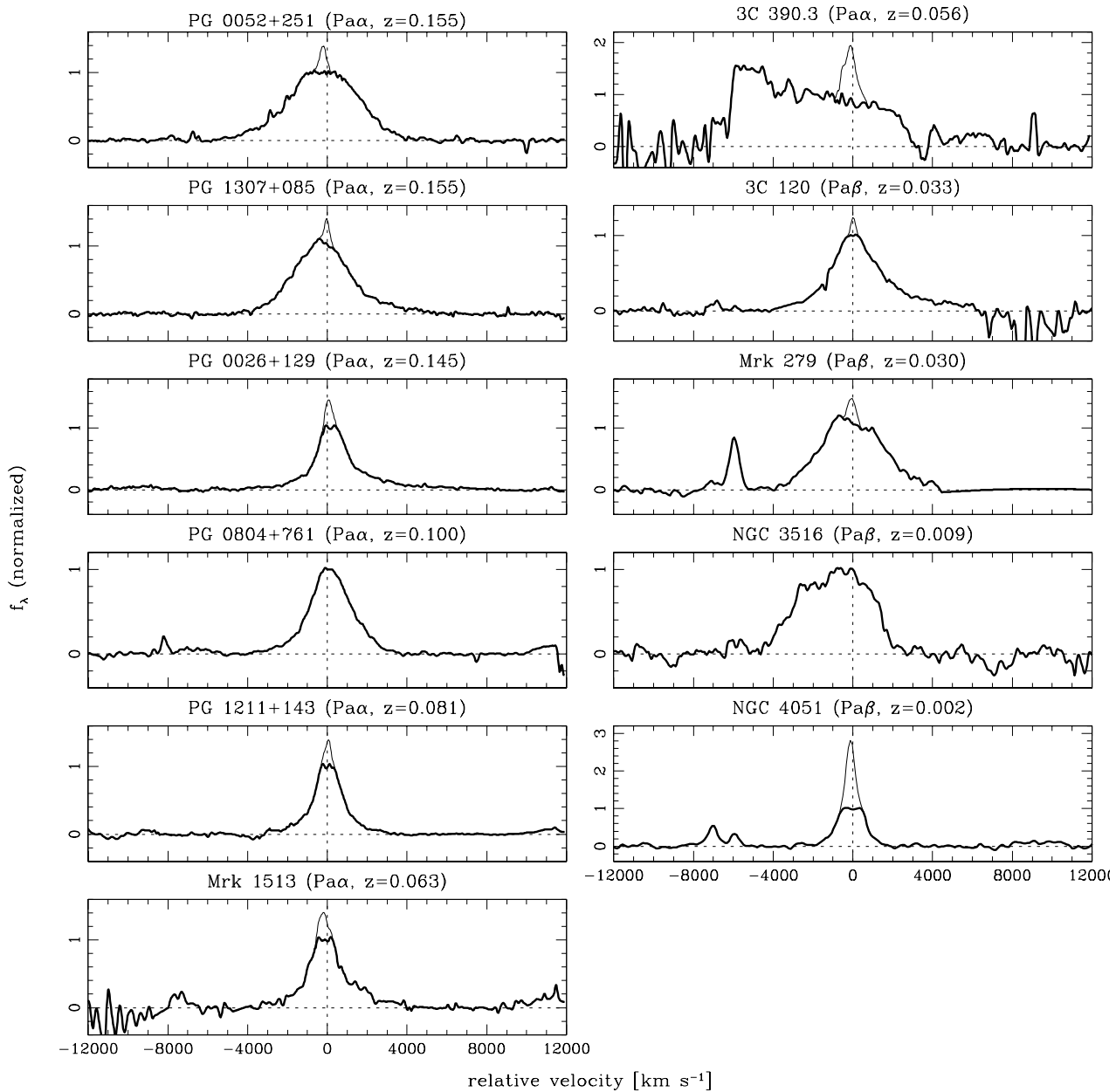}
}
\caption{\label{profiles} Pa$\alpha$ or Pa$\beta$ broad-emission line
  profiles in velocity space relative to the expected rest frame
  wavelength (thick lines). The profiles have been
  continuum-subtracted and normalized to the same peak intensity (of
  the broad component). The subtracted narrow components are also
  shown (thin lines).}
\end{figure*}

In Figs. \ref{mbhfwhm} and \ref{mbhdisp} (left panels) we show the
relation between the black hole mass derived from optical
reverberation mapping campaigns and the near-IR virial product. We
have calculated the latter using for the line width the full width at
half maximum (FWHM; Fig. \ref{mbhfwhm}) and the line dispersion
($\sigma_{\rm line}$; Fig. \ref{mbhdisp}) of the Pa$\alpha$ or
Pa$\beta$ broad component, whichever line had the higher
signal-to-noise ratio. Furthermore, we have assumed that the $R$--$L$
relationship has a logarithmic slope of 0.5, since our results in
Section \ref{radiuslum} were consistent with this value. We have not
plotted here the source Mrk~590, since it was found to have weak and
noisy broad emission lines. A crucial step towards isolating the
intrinsic broad-line profile is the subtraction of the narrow-line
component. The narrow-line profile appears inflected, i.e., the
transition point between the broad and narrow components is obvious,
in roughly half of our sample (14/26 sources; see also Table
\ref{sample} and Fig. \ref{profiles}). Four sources (PG~0844$+$349,
PG~0804$+$761, Ark~120, and NGC~3516) clearly lack a Paschen
narrow-line component, since their profiles have a broad top. However,
in the remainder (8/26 sources) the transition between the broad and
narrow components is not perceptible. As discussed in \citet{L08a}, in
these cases we have estimated the contribution of the narrow component
to the total profile by fitting to its top part a Gaussian with FWHM
equal to that of the narrow emission line \OIII~$\lambda 5007$. In the
case of the additional sample observed with GNIRS, for which we do not
have contemporaneous optical spectroscopy, we have used instead the
near-IR narrow emission line \SIII~$\lambda 9531$. This method assumes
that the FWHM of \OIII~$\lambda 5007$ or \SIII~$\lambda 9531$ is
representative of the narrow emission line region and subtracts the
largest possible flux contribution from this region.

As explained by \citet{Pet04}, both measures of the line width have
their advantages and disadvantages. Whereas the FWHM value can be
measured directly in the spectrum and is less sensitive to line
blending, the line dispersion is well-defined for arbitrary line
profiles and is less sensitive to the presence of even strong
narrow-line components. In particular for the latter reason, the black
hole masses from optical reverberation campaigns are always calculated
using the line dispersion, whereby this quantity is measured in the
rms spectrum, i.e., in the variable part of the spectrum.

\begin{figure}
\centerline{
\includegraphics[clip=true,bb=15 15 600 420,angle=-90,scale=0.5]{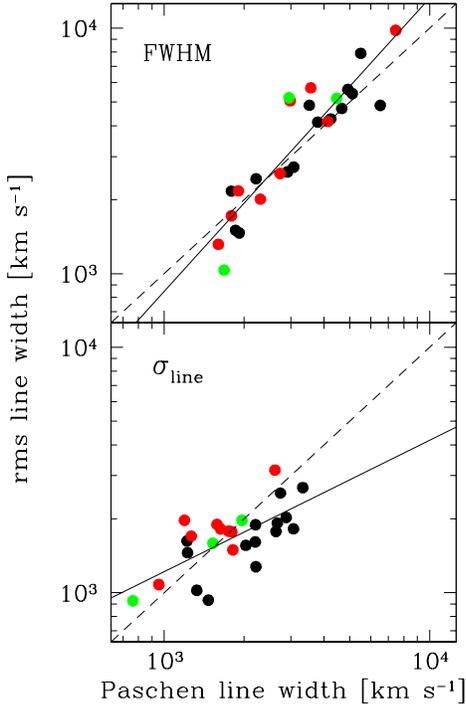}
}
\caption{\label{width} The width of the H$\beta$ broad emission line
  as measured in the rms spectrum versus that of the Pa$\alpha$ or
  Pa$\beta$ broad component, whichever line had the higher
  signal-to-noise ratio. The line width is measured as the full width
  at half maximum (FWHM; upper panel) and the line dispersion (lower
  panel). Symbols are as in Fig. \ref{lag}. The solid lines show the
  observed correlations and the dashed lines mark the locus of
  equality.}
\end{figure}

We have used the C routine MPFIT \citep[version 1.1;][]{mpfit}, which
solves the least-squares problem with the Levenberg-Marquardt
technique, to fit both correlations. The best-fit lines (solid lines
in Figs. \ref{mbhfwhm} and \ref{mbhdisp}) are:

\begin{eqnarray}
\label{mbhfwhmrel}
\log M_{\rm BH} = (0.88\pm0.04) \cdot (2 \log {\rm FWHM} + 0.5 \log \nu L_{1\mu m}) \nonumber \\
- (17.39\pm1.02)
\end{eqnarray}

\noindent
and

\begin{eqnarray}
\label{mbhsigmarel}
\log M_{\rm BH} = (0.98\pm0.04) \cdot (2 \log \sigma_{\rm line} + 0.5 \log \nu L_{1\mu m}) \nonumber \\
- (20.02\pm1.11),
\end{eqnarray}

\noindent
where $M_{\rm BH}$ is in solar masses, FWHM and $\sigma_{\rm line}$
are in km~s$^{-1}$ and $\nu L_{1\mu m}$ is in erg~s$^{-1}$. In both
cases the correlation slopes are close to one, which suggests that the
line width of an {\it unblended} broad emission line in a single-epoch
spectrum is a suitable proxy for the line width measured in an rms
spectrum. In order to verify this claim, we have plotted in
Fig. \ref{width} the line width of the H$\beta$ broad emission line as
measured in the rms spectrum versus that of the Pa$\alpha$ and
Pa$\beta$ broad components for both cases, i.e., using the FWHM (upper
panel) and the line dispersion (lower panel). In the case of FWHM, we
find a correlation slope consistent with unity of $1.2\pm0.1$,
however, the line dispersion in rms spectra appears to be smaller than
that in single-epoch near-IR spectra (correlation slope of
$0.5\pm0.1$). Since the line dispersion strongly depends on the total
flux of the emission line, this result could be explained if the broad
wings vary less than the core of the emission line since they are
composed of optically thin gas \citep{Shields95}.

In order to estimate black hole masses from single-epoch near-IR
spectra one then has two options: 1. directly applying the
relationship between black hole mass and near-IR virial product
(eqs. (\ref{mbhfwhmrel}) and (\ref{mbhsigmarel})); and 2. applying the
near-IR $R$--$L$ relationship together with the virial theorem
(eq. (\ref{virial})). In Figs. \ref{mbhfwhm} and \ref{mbhdisp} (right
panels) we show the results for the latter option using scaling factor
values of $f=1.4$ and 5.5 in the case of FWHM and line dispersion,
respectively \citep{Onken04}. Clearly, both measures of the line width
give results consistent with the reverberation-based black hole
masses. But we note that in the case of the line dispersion the second
(but not the first) option is expected to overestimate the black hole
mass (by a factor of $\la 2$) given the relationship between the
H$\beta$ rms line width and that of the Pa$\alpha$ and Pa$\beta$ broad
components presented in Fig. \ref{width}.

\section{Low-luminosity AGN} \label{lowlum}

\begin{figure}
\centerline{
\includegraphics[scale=0.45]{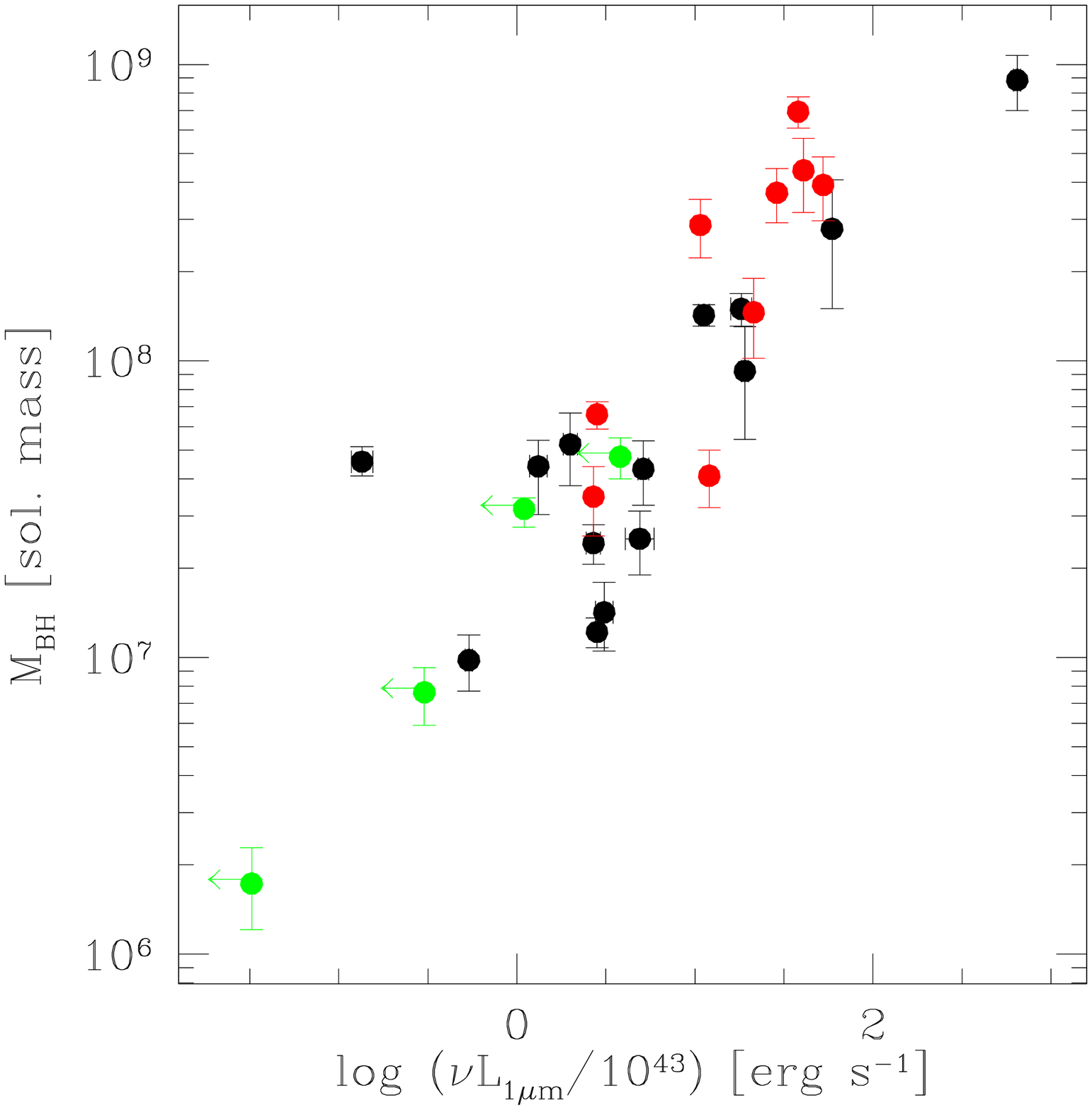}
}
\caption{\label{masslum} Black hole mass determined from optical
  reverberation campaigns versus the continuum luminosity at 1
  $\mu$m. Symbols are as in Fig. \ref{lag}.}
\end{figure}

Four AGN in our sample, namely, Mrk~590, NGC~3227, NGC~3516 and
NGC~4051, were found to be in a very low luminosity state with a
continuum SED dominated by host galaxy emission (see
Fig. \ref{SEDgal}, top panels, and \citet{L11a}, their
Fig. 6). Therefore, in these cases the observed 1~$\mu$m continuum
luminosity represents only an upper limit to the AGN value. As
Fig. \ref{lag} shows, these sources (shown as green filled circles)
populate indeed the low end of the near-IR luminosity distribution of
our sample. However, they follow not only the near-IR $R$--$L$
relationship remarkably well, but also the so-called black hole
mass-luminosity ($M$-$L$) relationship \citep{Kaspi00, Pet04}, which
we plot in Fig. \ref{masslum} for the near-IR. Therefore, the AGN
1~$\mu$m continuum luminosity of these sources cannot be significantly
(more than a factor of $\sim 2-3$) lower than that of the host galaxy.

Can and should black hole masses be estimated for AGN with 1~$\mu$m
continuum luminosities dominated by host-galaxy starlight? The near-IR
virial products and near-IR black hole mass values calculated for the
four host-galaxy dominated sources in our sample, although only upper
limits, follow well the general relationships (see Figs. \ref{mbhfwhm}
and \ref{mbhdisp}). But we also observe that the SEDs of all these
sources are dominated by the hot dust component at wavelengths
$\lambda \ga 1.5$~$\mu$m, i.e., in the near-IR H and K bands. 

\section{Summary and conclusions}

We have presented an improved calibration of our near-IR relationship
for estimating black hole masses by increasing the sample of
reverberation-mapped AGN with high-quality near-IR spectroscopy from
14 to 23 sources. The additional observations were obtained at the
Gemini North observatory with the Gemini Near-Infrared Spectrograph
(GNIRS). Our main results can be summarized as follows.

\vspace*{0.2cm}

(i) The near-IR radius -- luminosity ($R$--$L$) relationship has in
most cases a best-fitting slope of $\sim 0.6\pm0.1$. This value is
slightly steeper than our previous result but consistent with a value
of 0.5, which is expected based on simple photoionization
arguments. The scatter in the relationship stays unchanged at a value
of $\sim 50\%$, indicating that it is intrinsic rather than
observational. The additional sample has improved the number
statistics in particular at the high luminosity end.

(ii) The black hole mass derived from optical reverberation mapping
campaigns correlates strongly with the near-IR virial product, i.e.,
the product between the integrated total 1~$\mu$m continuum luminosity
and the width of the unblended Paschen hydrogen broad emission lines
Pa$\alpha$ and Pa$\beta$. The correlation slope is $\approx 1$, which
suggests that the line width of an {\it unblended} broad emission line
in a single-epoch spectrum is a suitable proxy for the line width
measured in an rms spectrum.

(iii) We have excluded four AGN in our sample from the calibration of
the near-IR relationship, since they were found to be in a very low
luminosity state and so to have a continuum emission dominated by host
galaxy starlight. Nevertheless, these sources still follow the near-IR
relationship remarkably well, indicating that it can be used to
reliably derive upper limits on the black hole masses of
low-luminosity AGN.

\vspace*{0.2cm}

In the future we plan to start a near-IR reverberation mapping
campaign, which will allow us to obtain time lags for the Paschen
broad emission lines and thereby black hole mass relationships fully
calibrated in the near-IR.

\section*{Acknowledgments}

We thank an anonymous reviewer for their thoughtful comments, which
improved the paper. We thank Chris Onken from the Australian Gemini
Office for his excellent user support. H. L. acknowledges financial
support by the European Union through the COFUND scheme. This work is
based on observations obtained at the Gemini Observatory, which is
operated by the Association of Universities for Research in Astronomy,
Inc., under a cooperative agreement with the NSF on behalf of the
Gemini partnership: the National Science Foundation (United States),
the Science and Technology Facilities Council (United Kingdom), the
National Research Council (Canada), CONICYT (Chile), the Australian
Research Council (Australia), Minist\'erio da Ci\^encia, Tecnologia e
Inova\c{c}\~ao (Brazil) and Ministerio de Ciencia, Tecnolog\'ia e
Innovaci\'on Productiva (Argentina). This research has made use of the
NASA/IPAC Extragalactic Database (NED), which is operated by the Jet
Propulsion Laboratory, California Institute of Technology, under
contract with the National Aeronautics Space Administration.

\bibliography{/Users/herminelandt/references}

\appendix
\section{Gemini Near-IR spectra}


\begin{figure*}
\centerline{
\includegraphics[scale=0.7, clip=true, bb= 35 235 580 700]{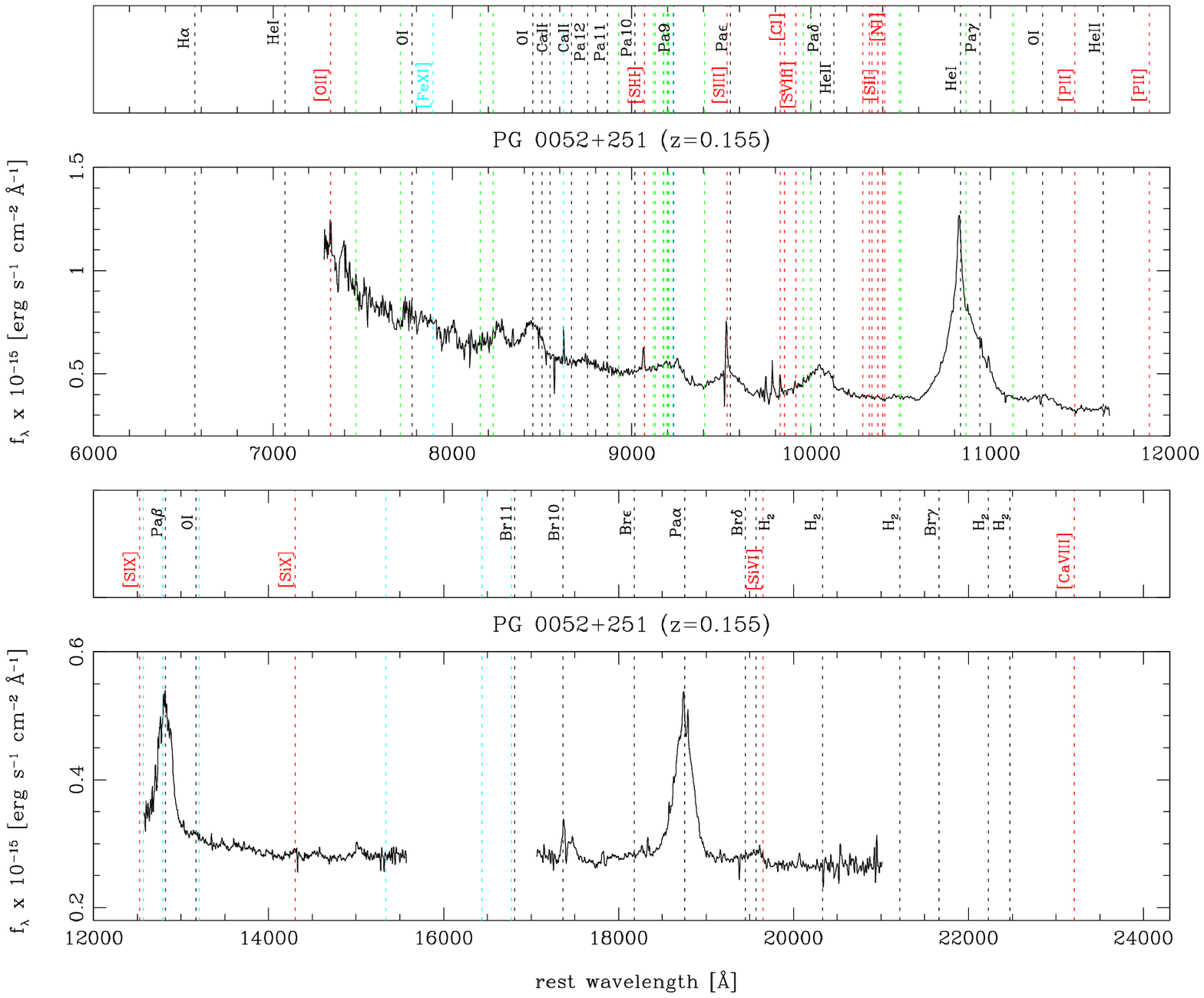}
}
\centerline{
\includegraphics[scale=0.7, clip=true, bb= 35 235 580 700]{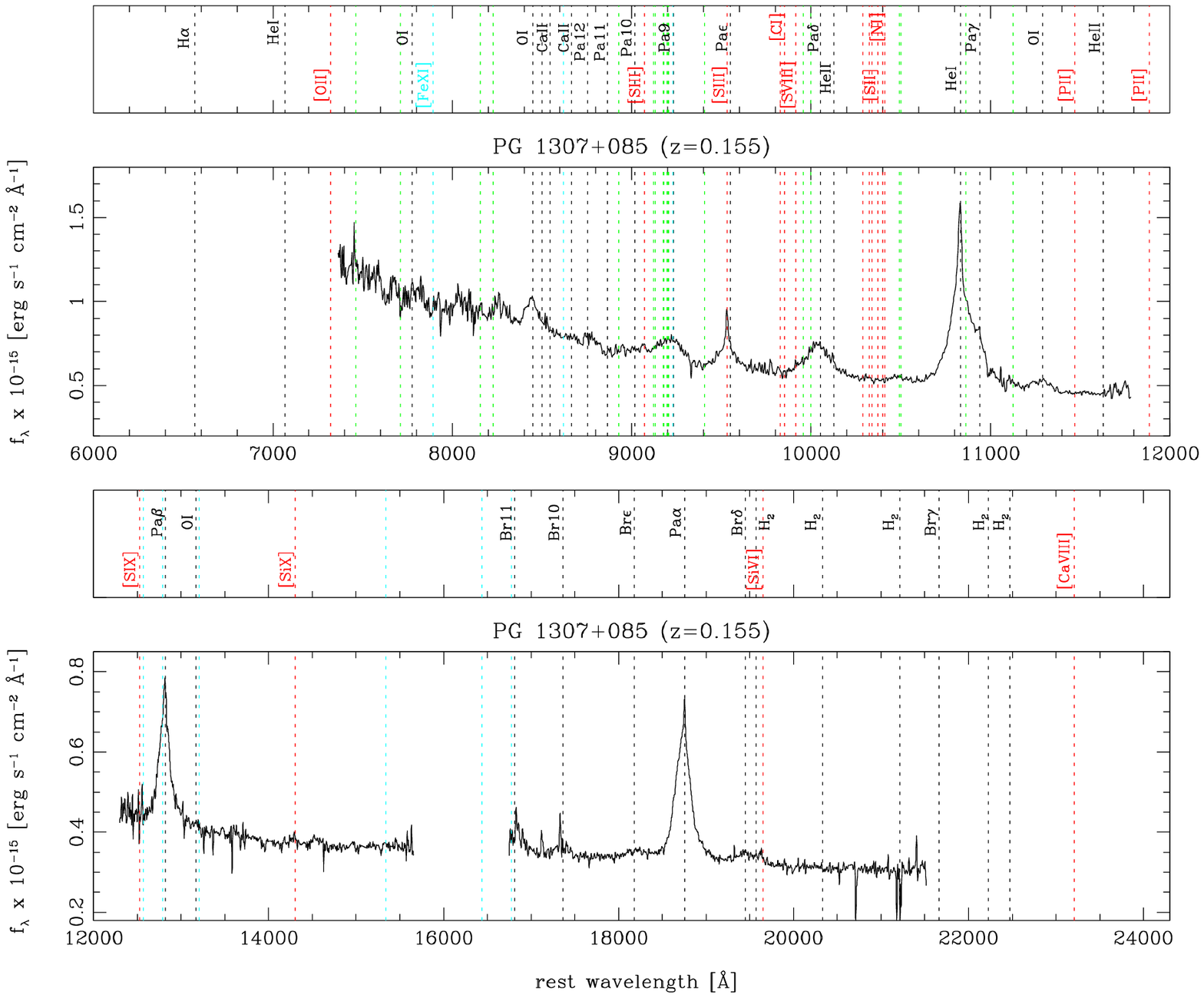}
}
\caption{\label{gnirsspec} Gemini GNIRS near-IR spectra shown as
  observed flux versus rest-frame wavelength. Emission lines listed in
  Table 4 of \citet{L08a} are marked by dotted lines and labeled;
  black: permitted transitions, green: permitted \FeII~multiplets (not
  labeled), red: forbidden transitions and cyan: forbidden transitions
  of iron (those of \FeIIf~not labeled).}
\end{figure*}

\setcounter{figure}{1}
\begin{figure*}
\centerline{
\includegraphics[scale=0.7, clip=true, bb= 35 235 580 700]{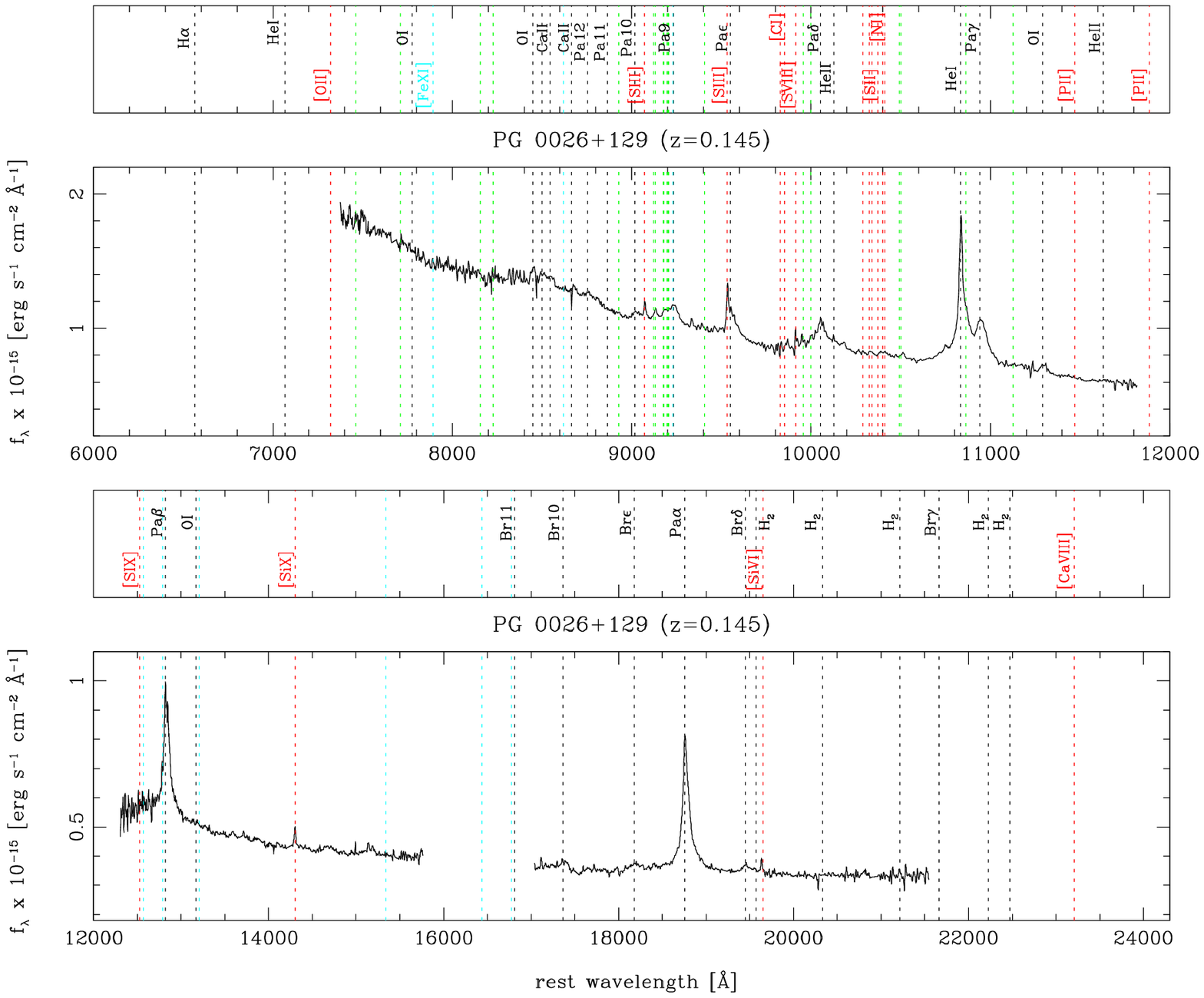}
}
\centerline{
\includegraphics[scale=0.7, clip=true, bb= 35 235 580 700]{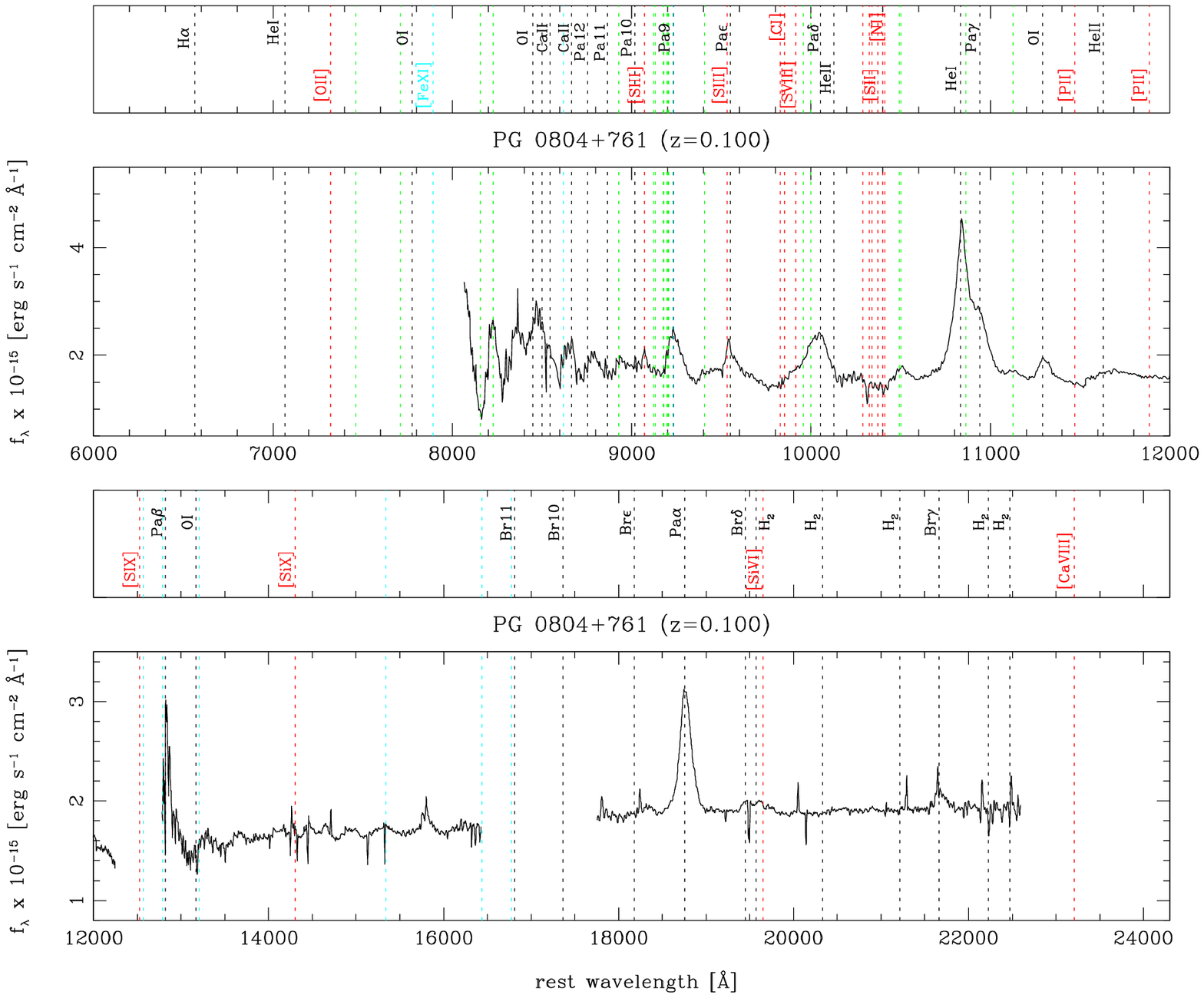}
}
\contcaption{}
\end{figure*}

\setcounter{figure}{1}
\begin{figure*}
\centerline{
\includegraphics[scale=0.7, clip=true, bb= 35 235 580 700]{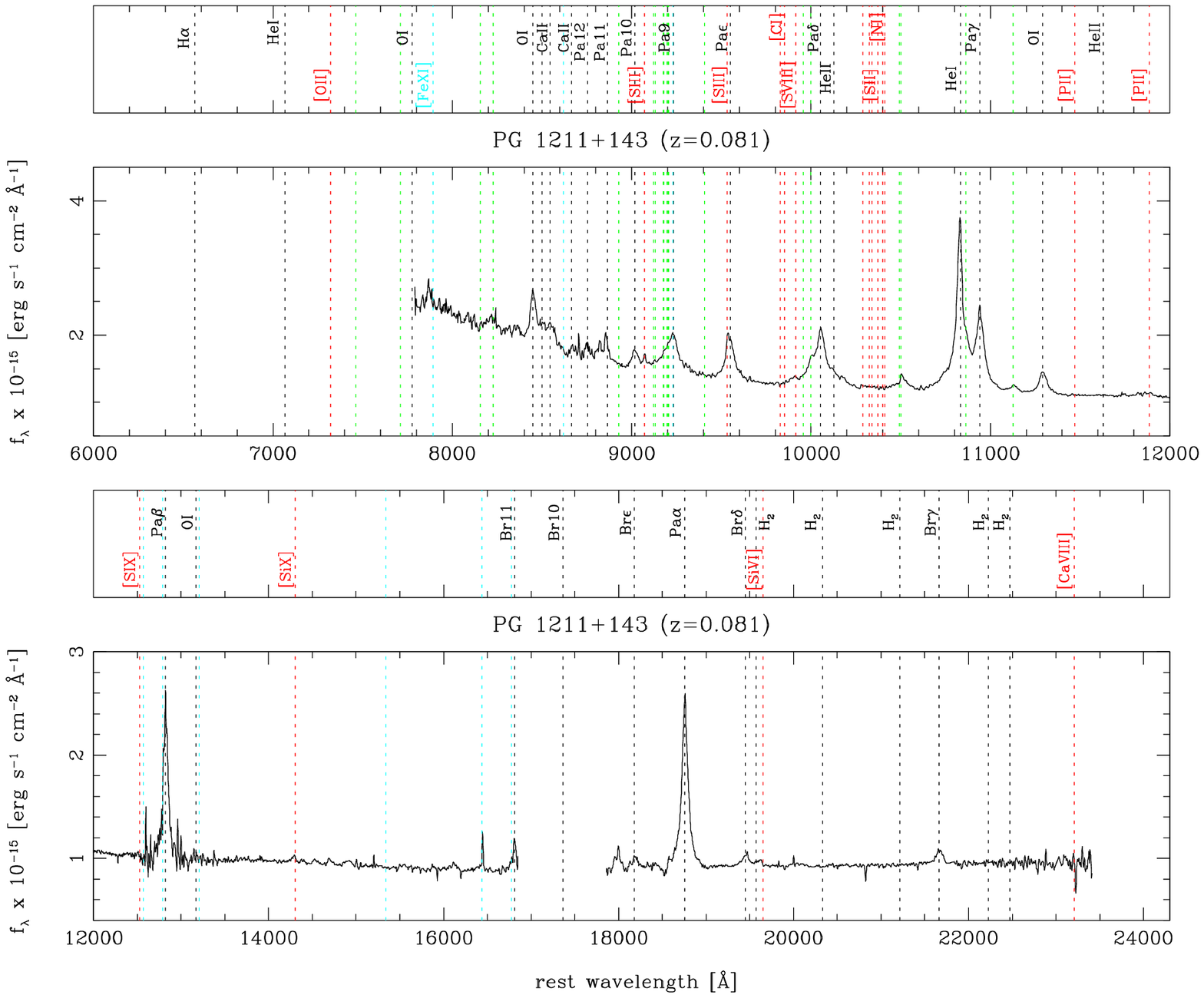}
}
\centerline{
\includegraphics[scale=0.7, clip=true, bb= 35 235 580 700]{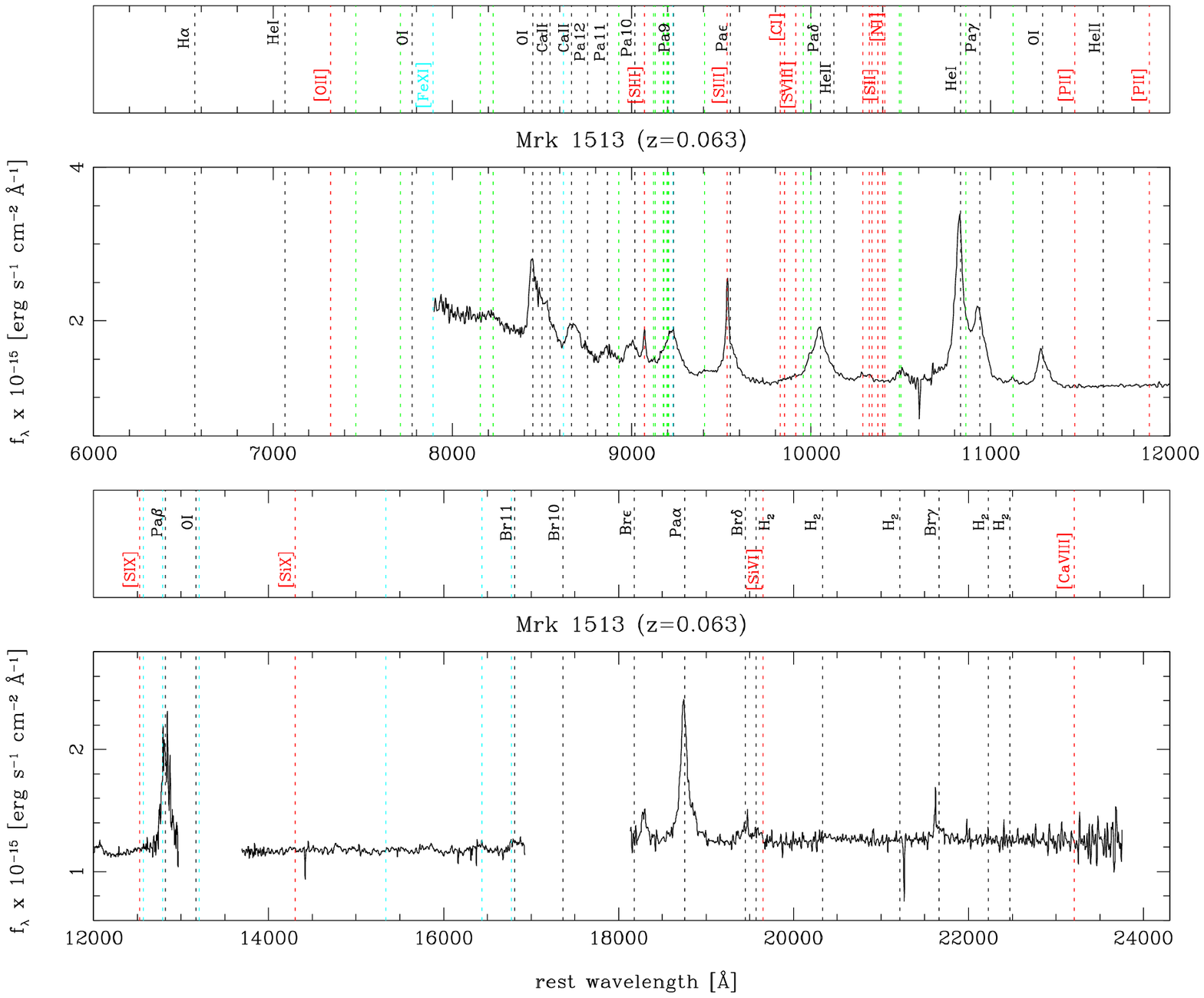}
}
\contcaption{}
\end{figure*}

\setcounter{figure}{1}
\begin{figure*}
\centerline{
\includegraphics[scale=0.7, clip=true, bb= 35 235 580 700]{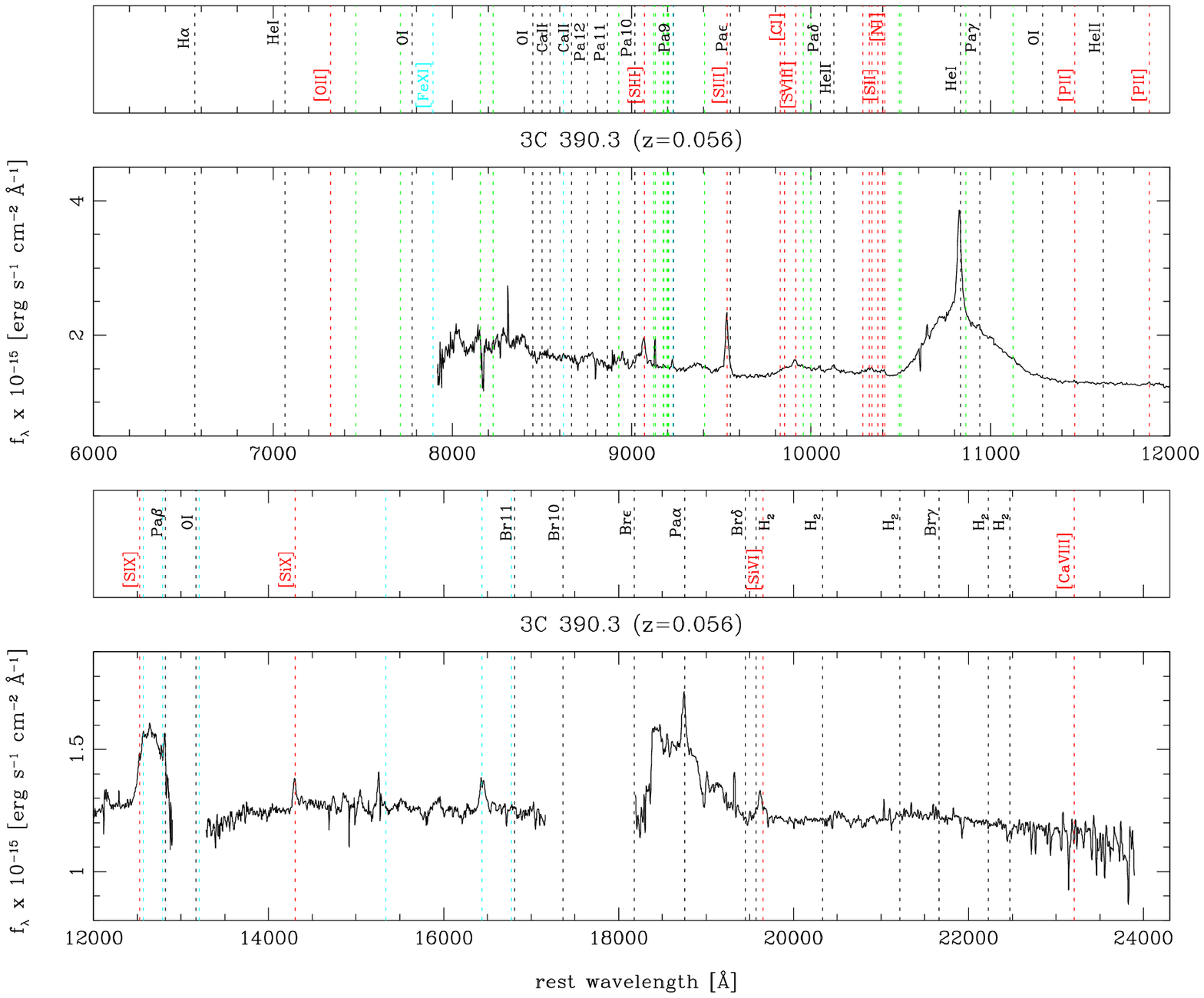}
}
\centerline{
\includegraphics[scale=0.7, clip=true, bb= 35 235 580 700]{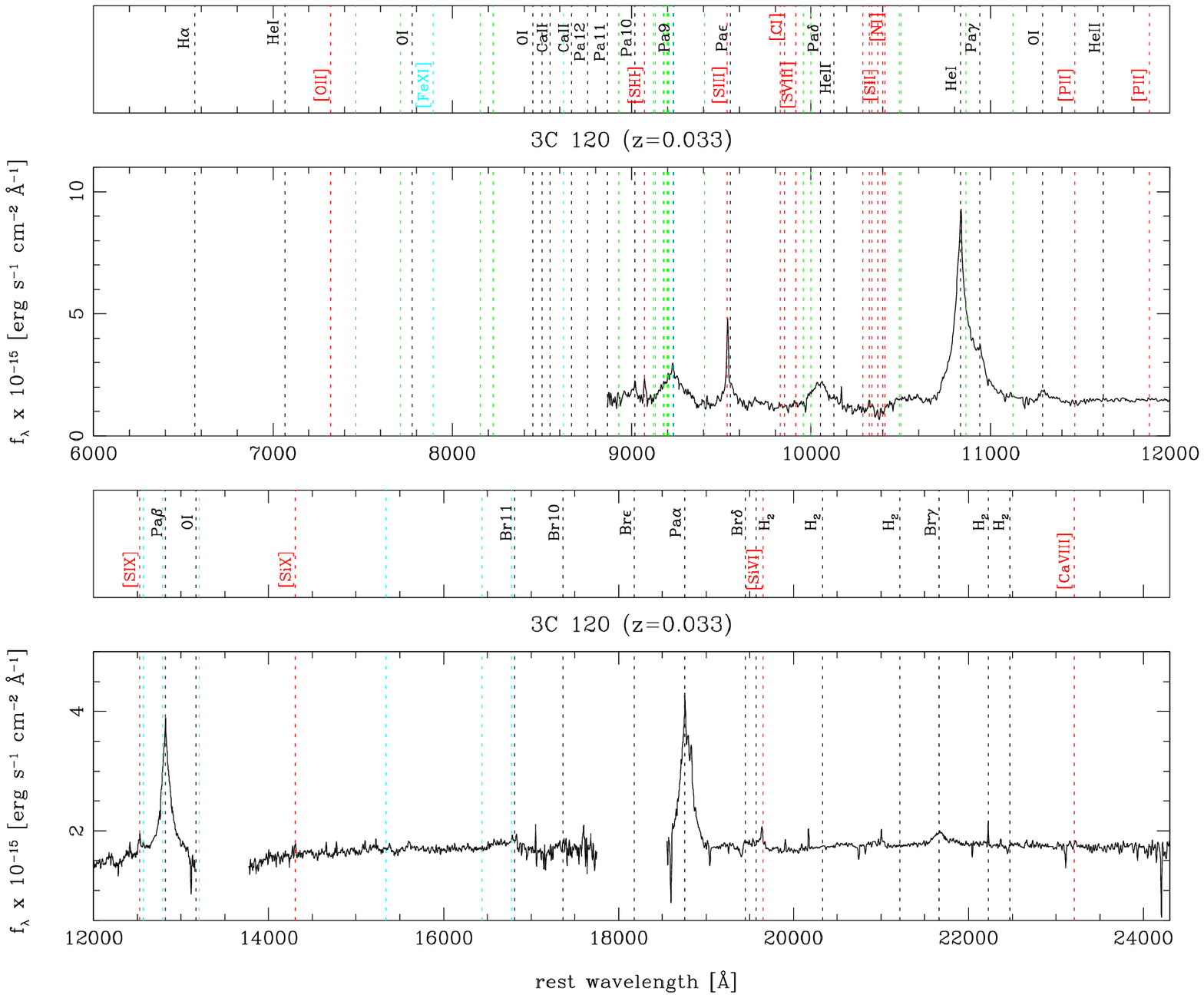}
}
\contcaption{}
\end{figure*}

\setcounter{figure}{1}
\begin{figure*}
\centerline{
\includegraphics[scale=0.7, clip=true, bb= 35 235 580 700]{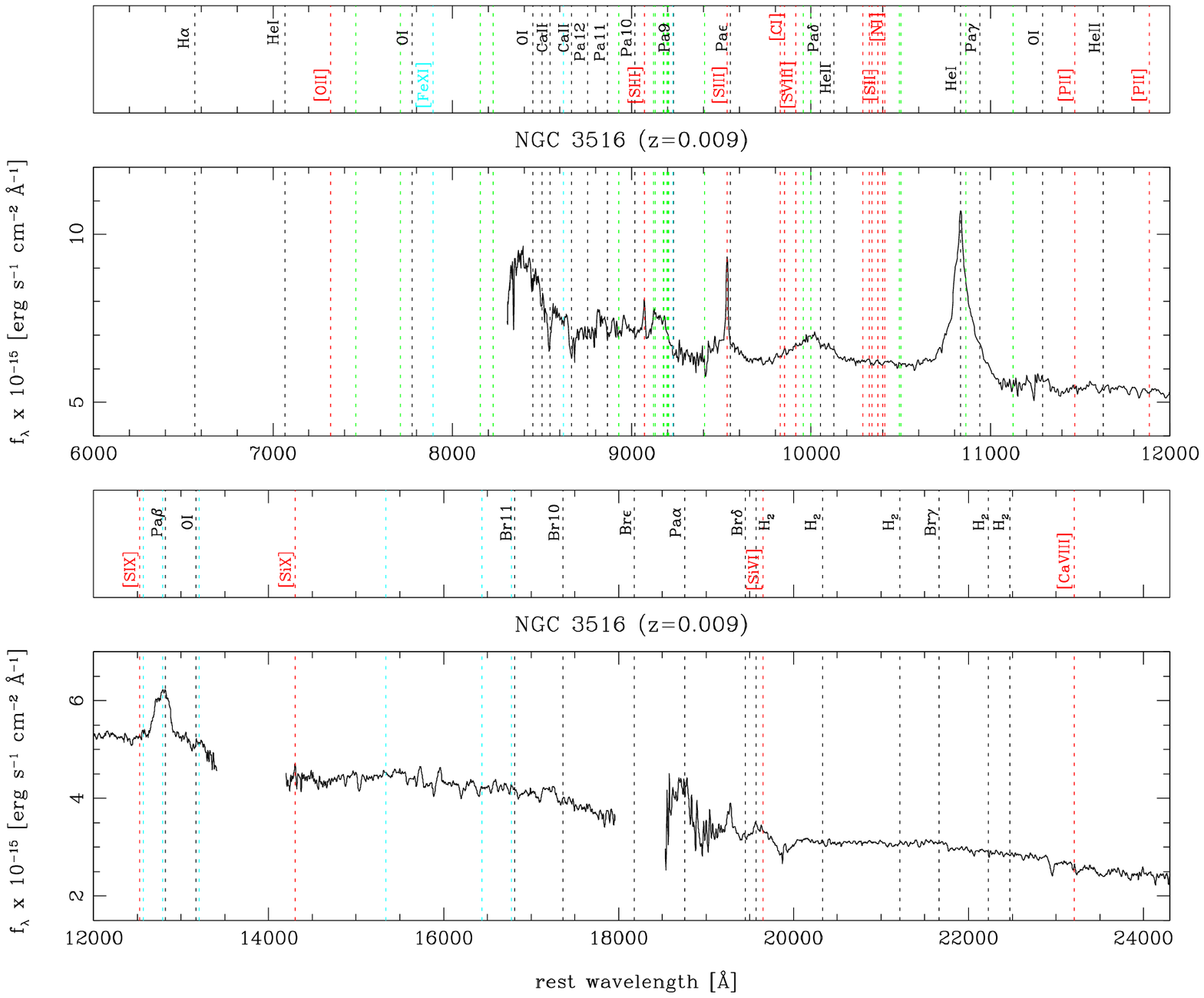}
}
\contcaption{}
\end{figure*}

\bsp
\label{lastpage}

\end{document}